\documentclass[twoside]{ilcws08}
\usepackage[latin1]{inputenc}
\usepackage[dvips]{graphicx,epsfig,color}
\usepackage{wrapfig,rotating}
\usepackage{amssymb,amsmath,array}

\usepackage{booktabs}
\usepackage{listings} 
\usepackage{subfigure}

\begin{document}
\title{
Implementation and Application of Kinematic Vertex Fitting in the 
Software Environment of ILD} 
\author{{\sl Fabian Moser, Wolfgang Waltenberger, Meinhard Regler} and {\sl Winfried Mitaroff}
\vspace{.3cm}\\
Austrian Academy of Sciences -- Institute of High Energy Physics \\
Nikolsdorfer Gasse 18, A-1050 Vienna, Austria, EU
}

\maketitle

\begin{abstract}
The vertex reconstruction toolkit RAVE has been extended by an option for the inclusion of kinematic constraints, and embedded into the ILD analysis framework Marlin. 
The new tools have been tested with an exemplary reconstruction of $WW$ and $ZZ$ decays. The presented results show the improvements achieved in precision of the fitted masses, and demonstrate the usage and functionality of the toolkit.
\end{abstract}

\section{Software elements}

\subsection{The Rave library}

The Rave library \cite{RaveHomepage} was created with the aim of avoiding repeated re-implementation of similar algorithms in every new reconstruction software, when specific modules can be formulated in an experiment-independent manner. This is certainly true for the modules doing the reconstruction of interaction vertices, e.g. of particle decays.

The core of the library is an implementation of algorithms for geometric vertex reconstruction which were developed for the CMS experiment at the Large Hadron Collider, augmented by an additional simple and stable interface. 
The same approach was taken for this work: the algorithms doing vertex reconstruction with kinematic constraints \cite{Kirill} were extracted from the CMSSW framework and implemented in the library.

\subsection{The MarlinRave plug-in}

A plug-in has been developed for the ILD analysis framework Marlin \cite{GaedeMarlin}, 
enabling users to access the geometric and kinematic vertex reconstruction capabilities 
of RAVE. Its name is MarlinRave, and it provides two new processors to the framework -- 
RaveVertexing and RaveKinematics -- one of which or both to be selected by the user.

\vspace{-0.3cm}
\subsubsection{The RaveVertexing processor}
\vspace{-0.2cm}
Geometric vertex reconstruction (by linear or adaptive filter algorithms) is accessed from Marlin through the RaveVertexing processor. 
The configuration of this processor requires an input collection containing objects of type \lstinline|EVENT::Track|. It takes a text string configuring the vertexing algorithm to be used together with its parameters, and the names of two output collections: one containing the fitted vertices, and the other containing the re-fitted (``smoothed'') tracks at their vertex.

\vspace{-0.3cm}
\subsubsection{The RaveKinematics processor}
\vspace{-0.2cm}
Its algorithm is based on a linear filter with Lagrangian multipliers for the constraints. 
Because the configuration interface of Marlin does not allow for arbitrary nesting, the RaveKinematics processor is designed to contain a flexible number of pre-defined kinematic decay topologies, thus facilitating the inclusion of new topologies as needed by the user. From the Marlin configuration file only the topology is chosen, whilst a text string allows the passing of specific parameters.

\section{Reconstruction of $W^{+}W^{-}$ decays}

\begin{wrapfigure}[38]{r}{0.5\columnwidth}
  \centerline{\subfigure[All $W$ masses as generated by Pythia.]{\label{fig:TrueMasses}\includegraphics[width=0.45\columnwidth]{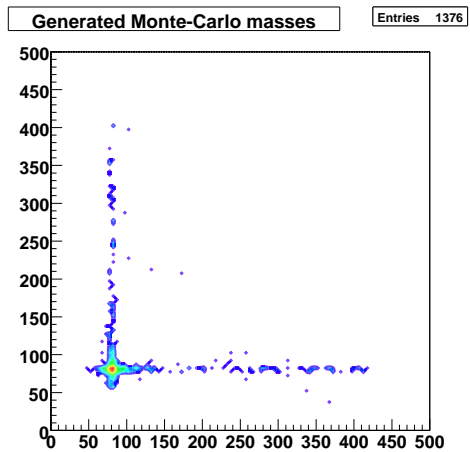}}}
  \centerline{\subfigure[The $W$ masses resulting from all possible combinations of jet association.]{\label{fig:AllAssociations}\includegraphics[width=0.45\columnwidth]{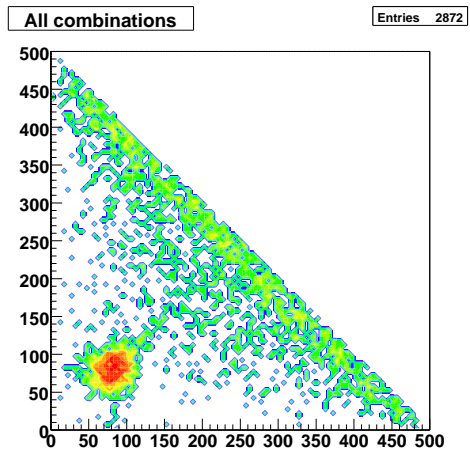}}}
  \caption{{\sl
  Scatter plots of the $W$ pair masses. The colours represent the number of entries in each bin.}}
\end{wrapfigure}

RAVE's kinematic capabilities were tested with a data sample consisting of 1376 events 
$e^{+}e^{-} \rightarrow W^{+}W^{-} \rightarrow 4$ jets at $\sqrt{s} = 500$~GeV, 
generated by Pythia \cite{Pythia6} and processed for the LDC 2007 layout \cite{MCRun}; 
the ``true'' $W$ masses are plotted in Figure~\ref{fig:TrueMasses}.
Events with a mass $| m_W - 80.32 | > 6.5$~GeV are eventually discarded, 
retaining 1008 events.

To model the expected performance of jet reconstruction in ILC experiments, the following Gaussian errors have been applied to all four generated jets:
\begin{align}
  &\sigma_E / E = 30 \% / \sqrt{E} \\
  &\sigma_\theta = \sigma_\phi \sin(\theta)  = 10\:\text{mrad} 
\end{align}
The direction resolution affects the $\theta$ (polar angle) and $\phi$ (azimuth) measurements. 

A known general problem with the use of jets as virtual measurements is due to their compositeness: only direction and energy, but not an absolute momentum, are well-defined. However,
an initialization of the jet's momentum with its energy, followed by an appropriate inflation of the associated error, gives satisfactory results.

The constraints applied are those of energy and 3-momentum conservation. They are explicitly written as (the subscripts identify the four jets):
\begin{align}
  E_1 + E_2 + E_3 + E_4 &= \sqrt{s} \\
  \vec{p}_1 + \vec{p}_2 + \vec{p}_3 + \vec{p}_4 &= \vec{0}
\end{align}

So far no distinction was made between the four jets: the applied kinematic constraints acted equally on all final states, and did not take into account which particles they could have originated from. 
Now one has the task of associating the four jets into two pairs (there are 3 possible combinations), with each pair originating from a $W$ boson decay.

A trivial strategy is to assume that each combination is valid, and to calculate the two $W$ masses accordingly. Figure~\ref{fig:AllAssociations} shows the results: each entry represents the $W$ pair masses corresponding to one combination of one event.\footnote{
About 5\% of the combinations yield unphysical mass values and do not enter the plot.}

A comparison of Figure~\ref{fig:AllAssociations} with \ref{fig:TrueMasses} suggests that,  as a first step, 
dropping combinations where both $W$ masses exceed a pre-defined cut limit would significantly improve the performance of the association. This cut is chosen at a value of 130~GeV.

\begin{wrapfigure}[22]{r}{0.5\columnwidth}
  \centerline{\includegraphics[width=0.45\columnwidth]{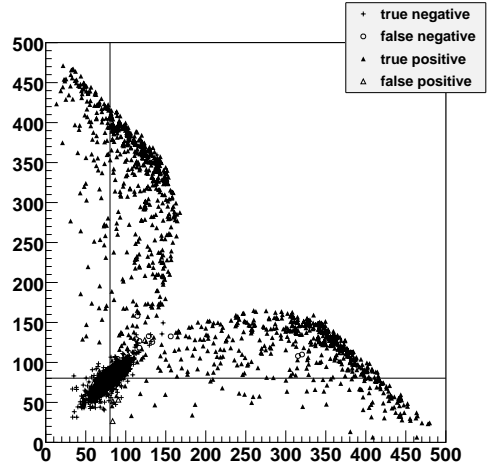}}
  \caption{{\sl 
  Scatter plot of the $W$ pair masses fitted with the similar-mass constraint; 
  selection of the best jet association by the fit's pseudo-$\chi^2$ probability, 
  applied to the reduced sample after the 130}~GeV {\sl cut.}}
  \label{fig:GaussLowCorrCut}
\end{wrapfigure}

For a second step, several options are feasible. A simple ``equal-mass hypothesis'' introduces in the kinematic fit, as an additional constraint, the requirement of the two fitted masses to be equal (because they both belong to $W$ bosons). However, such a requirement would strongly favour combinations along the $45^o$ diagonal over those parallel to the axes, in contradiction to the true distribution of the $W$ pair masses as shown in Figure~\ref{fig:TrueMasses}.

The most obvious improvement of such a hypothesis would be to model a selection requirement from two uncorrelated Breit-Wigner (BW) distributions.\footnote{
More precisely: from $\sigma(s) = \int\limits_0^s \mathrm{d}s_A 
\int\limits_0^{s - s_A} \mathrm{d}s_B \rho(s_A) \rho(s_B)$ 
with $\rho(s_i) \propto \mathrm{BW} (M_W, \Gamma_W, s_i)$ \cite{Denner}.}
But that is not possible in real-world scenarios, because the position parameter of the expected distribution is exactly the (unknown) value to be determined by this kinematic vertex fit. 
Therefore, a compromise is suggested between this idea and the ``equal-mass hypothesis'' above
\cite{MoserThesis}:

The two $W$ masses are known to not being equal, however, they are still picked from the same distribution.
The likelihood of such a configuration is given by 
\begin{equation}
  \label{equ:MassLikelihood}
  \mathrm{L}(\vec{\alpha}_c | \vec{\alpha}_m, \bar{m}) = \mathrm{L}(\vec{\alpha}_c | \vec{\alpha}_m) 
  \cdot \mathrm{L}(m_1(\vec{\alpha}_m), m_2(\vec{\alpha}_m) | \bar{m})
\end{equation}

Here, the first term holds the results of the kinematic vertex fit (parameters $\vec{\alpha_c}$ and covariances $\mathbf{V}_c$). The second term represents the new ``similar-mass hypothesis''; it models the distribution of the two masses dependent on the position parameter $\bar{m}$, and it can be split into two symmetric terms if the masses are approximately uncorrelated. Although any p.d.f. can be used within the second term, a Gaussian model is certainly the easiest choice. Then, the objective function for the new fit can be written as 
\begin{equation}
  \mathrm{M}(\vec{\alpha}_m, \bar{m}) = 
  \left[ \vec{\alpha}_m - \vec{\alpha}_c \right]^\mathrm{T} \mathbf{V}_c^{-1}
  \left[ \vec{\alpha}_m - \vec{\alpha}_c \right] +
  \sum\limits_{i=1}^2 \frac{\left( m_i\left(\vec{\alpha}_m\right) - \bar{m} \right)^2}{\sigma_t^2}
\end{equation}
where $\sigma_t \approx 9$~GeV is a pre-defined scale (obtained by integration over the BW).

The objective function M is minimized w.r.t. $\vec{\alpha}_m$ and $\bar{m}$ and yields a 
pseudo-$\chi^2$, the probability of which is used as the new second step selection criterion for finding the best jet association (out of the 3 possible combinations).

\begin{wrapfigure}[16]{r}{0.6\columnwidth}
  \centerline{\includegraphics[width=0.55\columnwidth]{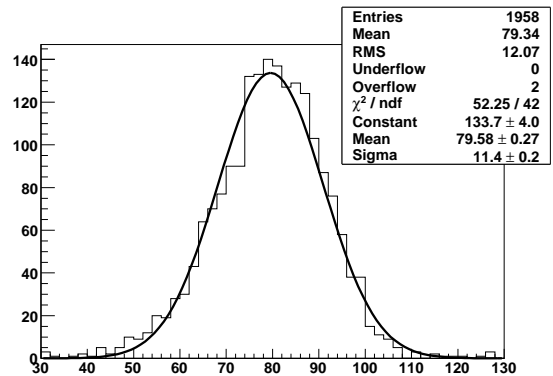}}
  \caption{{\sl
  Reconstructed $W$ masses after kinematic vertex fitting with the similar-mass constraint.}}
  \label{fig:WWReco}
\end{wrapfigure}


This strategy, after application of the first step (130~GeV cut mentioned above), results in 
979 events; the $W$ pair masses are shown in Figure~\ref{fig:GaussLowCorrCut}. 

The association performance of this method is characterized by type 1 and 2 errors 
of $1.6$\,\% and $0.9$\,\%, respectively.

Figure~\ref{fig:WWReco} shows the reconstructed $W$ masses after kinematic vertex fitting, 
including the similar-mass constraint.\footnote{
Note that the four kinematic constraints (equ. 3 and 4) and the similar-mass constraint (equ. 6) 
together do not suffice to account for the redundant degrees of freedom.}

A Gaussian fit over the histogram range $30 \ldots 130$~GeV yields for mean and standard deviation:
$\mu(m_W) = 79.58 \pm 0.27$~GeV and $\sigma(m_W) = 11.4 \pm 0.2$~GeV.

Scaled to 664k events expected in 4 years at the ILC 
(500~fb$^{-1}$, 4-jet efficiency $\approx 40\%$) \cite{MoserThesis}, 
an accuracy of 0.014~GeV in determining the $W$ mass may be achieved.

\section{Reconstruction of $Z^{o}Z^{o}$ decays}


The other exemplary reconstruction was performed on a data sample of 994 Pythia events 
$e^{+}e^{-} \rightarrow Z^{o}Z^{o} \rightarrow 4$ jets at $\sqrt{s} = 250$~GeV, 
and processed for the LDCprime 2008 layout \cite{MCRunZZ}. 
In contrast to the $WW$ data of chapter 2, simulation included initial state radiation; 
at present, this cannot be accounted for in our kinematic reconstruction. Therefore, 
events with a total $ZZ$ energy $< 249$~GeV are discarded, retaining 531 events.

The $ZZ$ data had the momenta of the jets correctly calculated by PFA, so they were used in this study. 
Apart from that, exactly the same strategies and methods as in chapter~2 have been applied.
A comparison (not shown) of the reconstructed $Z$ masses with and without kinematic constraints, 
respectively, reflects significant improvement. 

%


\begin{footnotesize}

\bibliographystyle{unsrt}
\bibliography{references}

\end{footnotesize}


\end{document}